\documentclass[12pt]{article}
\usepackage{latexsym}
\usepackage[english]{babel}
\usepackage{amsmath}
\usepackage{graphics}
\usepackage{epsfig}
\usepackage{epstopdf}
\usepackage{amsmath}
\usepackage{array}
\begin{document}
\begin{center}
{\LARGE Bayesian analysis of bulk viscous matter dominated
universe \\[0.2in]}
{ Athira Sasidharan$^*$, N D Jerin Mohan$^*$, Moncy V John$^+$ and Titus K Mathew$^*$ \\
$^*$Department of
Physics, Cochin University of Science and Technology, India,\\$^+$School of Pure and Applied Physics, Mahatma Gandhi University, India,\\e-mail:athirasnair91@cusat.ac.in, jerinmohandk@cusat.ac.in, moncyjohn@yahoo.co.uk, titus@cusat.ac.in }
\end{center}

\date{\today}% It is always \today, today,
             %  but any date may be explicitly specified
%\maketitle

\abstract{In our previous works, we have analyzed the evolution of bulk viscous matter dominated universe with a more general form for bulk
viscous coefficient,
$\zeta=\zeta_{0}+\zeta_{1}\frac{\dot{a}}{a}+\zeta_{2}\frac{\ddot{a}}{\dot{a}}$ and also carried out the dynamical system analysis. We found that the model reasonably describes the evolution of the universe 
if the viscous coefficient is a constant. In the present work we are contrasting this model with the standard $\Lambda$CDM model of the universe using the Bayesian method. We have shown that, even though the viscous model gives a reasonable back ground evolution of the universe, the Bayes factor of the model indicates that, it is not so superior over the $\Lambda$CDM model, but have a slight advantage over it.}

\section{Introduction}
\label{intro}
Many observations lead to the conclusion that the present universe is accelerating \cite{Riess1,Perl1,Bennet1,Tegmark1,Seljak,Komatsu1}. The reason for this acceleration was attributed to the dominant presence of a new cosmic component called dark energy.  The $\Lambda$CDM model came out as the most successful one for explaining this late time acceleration of the universe. In this model the cosmological constant is being considered as the dark energy. But the model is plagued with severe drawbacks. The foremost among them is the cosmological constant problem and is about the discrepancy between the observed and predicted values of the dark energy density, which is of the order of 120.
The other is the coincidence problem, the mysterious coincidence between the energy densities of the dark energy and dark matter component during the current epoch of the universe in spite of their completely different evolution history. This motivates a large class of models with varying dark energy density \cite{Weinberg1,fujii,carroll,ford,Liddle,chiba1,Picon2000,Copeland1}. Perfect fluid models like Chaplygin gas model \cite{kamen1,Bento2002} would be an alternate suggestion, due to their ability to explain both the deceleration and late acceleration by a single cosmic component, which thus effectively leads to a unification of the dark matter and dark energy sectors. There were also attempts to study this phenomenon by modifying the geometry part of the gravity theories, like $f(R)$ gravity ~\cite{capo1,Thomas2010,Nojiri}, $f(T)$\ gravity
~\cite{ferraro1,Ratbay2011}, Gauss-Bonnet theory ~\cite{nojiri},
Lovelock gravity ~\cite{pad2}, Horava-Lifshitz gravity
~\cite{horava1}, scalar-tensor theories ~\cite{amendola1},
braneworld models ~\cite{dvali1} etc. 

As in the case of the Chaplygin gas model, another possibility of the unified description of both dark energy and dark matter arises in the dissipative fluid models. 
It has been shown that the early inflationary period of the universe can be due to the presence of an imperfect fluid with bulk viscosity\cite{pad,waga,Cheng,Zimdahl,Gron}. This motivates the study of the disspiative cosmologies in the context of the late acceleration of the universe \cite{fabris1,li1,Hiplito1,av1,av2}. In \cite{li1}, by considering a single cosmic component, which is the dark matter with bulk viscosity, $\zeta(\rho)=\alpha \rho^m$ with $\alpha$ and $m$
being constants, the authors have shown that, the universe can make a transition from a decelerating phase to a late accelerating phase and ultimately to a de Sitter epoch. Inspite of this good background evolution, the model have come across with some negative aspects while analysing the structure formation. For instance, in reference \cite{li1} with $\zeta=\alpha \rho^{-0.4}$ the authors have shown that the density perturbation would rapidly be damped out,  which adversely affect mainly the CMBR. It may be due to the power factor of the density $-0.4$, which was obtained by constraining the model with old supernovae luminosity data by Riess and others\cite{Riess1}. At around the same time, in reference \cite{av1}, the authors have considered a constant bulk viscous dark matter dominated universe, with $0<\zeta<3$ and predicts that the universe began with a Big-Bang, followed by a decelerated expansion epoch and later transition into an accelerated epoch. Later these authors\cite{av2} extended their model by taking varying bulk viscosity of the form $\zeta=\zeta_0+\zeta_1 H$ and found that it shows a background evolution close to that of the standard $\Lambda$CDM model. In \cite{Anand}, the authors have shown that the data from Planck CMB observation and different LSS observations prefer small but non-zero amount of viscosity in cold dark matter fluid.

In \cite{Athira1}, we have analyzed the evolution of bulk viscous matter dominated universe with a more general form for bulk
viscous coefficient,
$\zeta=\zeta_{0}+\zeta_{1}\frac{\dot{a}}{a}+\zeta_{2}\frac{\ddot{a}}{\dot{a}}$, such that the viscosity depends on both the velocity and acceleration of the expansion, 
where $a$ is the scale factor of expansion of the universe. We have studied the model by contrasting it 
with Union compilation of supernovae data and found that it predicts the transition to 
the late accelerating phase. A similar work has been done in reference \cite{Avelino}, however, in constraining the parameters ($\zeta_{0}$, $\zeta_{1}$, $\zeta_{2}$), the authors fixed either $\zeta_{1}$ or $\zeta_{2}$ as zero while evaluating the other. In a later work \cite{Athira2} we found that  a general evolution of the universe with  
all the conventional phases (radiation epoch, matter dominated epoch and later acceleration) can be described without any causality violation only when the bulk viscous coefficient is a constant, $\zeta=\zeta_{0}$.

In the present paper we concentrate on the Bayesian analysis of the bulk viscous model corresponding to the different forms of the bulk viscous coefficient. We intend to compare these models in general with the standard $\Lambda$CDM model and also among themselves. Bayesian model comparison method is commonly used in the context of cosmological model selection \cite{Liddle1,Liddle2,Davis,Kurek,Daniel,Santos,Moncy1}. More details regarding Bayesian model comparison are discussed in the section \ref{sec:bayes}.

The paper is organized as follows. In Section \ref{sec:bayes}, we discuss the basic details regarding the Bayesian analysis. In section \ref{sec:flrw universe}, we introduce the bulk viscous model of the universe. In section \ref{sec:Analysis}, we extract the bulk viscous parameters corresponding to different cases of our model and discuss the results of the Bayesian analysis of the viscous model and finally, in section \ref{Sec:Conclusion}, we present our conclusion.

\section{Bayesian model comparison}
\label{sec:bayes}
Various models have been proposed to interpret the cosmological observational data which eventually add to our understanding of the evolution of the universe. So there exist, in fact many models explaining the expected evolution of the universe. Contrasting these models among themselves to select the better ones is essential for understanding the true evolution of the universe. Bayesian statistical approach\cite{Moncy1,Drell,Moncy2} is an effective tool to compare the new models with the standard $\Lambda$CDM model and also among themselves. The basic approach of this method is originated from the theory of random variables. In general, the relative merit of a random variable can be obtained by calculating the basic probability of it among the ensemble of values obtained theoretically or through repeated observations. But in cosmology repeated observations are virtually impossible. 
Here, what one can often do is to form hypothesis or a theory. For making the decision regarding the viability of such a proposed theory one have to assign certain probability to it in contrast to other theories existing for the same purpose. It is in this stage the Bayesian theory help us, so as to assign probability for a certain hypothesis by considering the observational data already available to us. Due to the acquisition of more  data, one can in fact adjust the plausibility of the hypothesis using Bayesian theorem.  This method  have been adopted by many in the past, for instance, Jaffe \cite{Jaffe} and Hobson et al. \cite{Hobson1} have analysed the relative merits of certain cosmological models. Also John and Narlikar\cite{Moncy1} have compared  a simple cosmological model with scale factor $a(t) \propto t$ with standard and inflationary models of the universe. In many models one does not have a prior knowledge about the model parameters for assigning the corresponding probability and in such cases one often starts with a flat prior for the parameter. 

According to Bayes's theorem\cite{Bayes}, the posterior probability $p(H_i|D,I)$ of a hypothesis $H_i,$ given the data $D$ and assuming any other background information $I$ to be true, is given as,
\begin{equation}
p(H_i|D,I)=\frac{p(H_i|I)p(D|H_i,I)}{p(D|I)}
\end{equation}
where $p(H_i|I)$ is the prior probability, i.e., the probability of $H_i$ given $I$ is true and $p(D|H_i,I)$ is the likelihood for the hypothesis $H_i$, which is the probability for obtaining the data D provided the hypothesis $H_i$ and $I$ are true. The factor $p(D|I)$ helps in normalization.

In Bayesian model comparison, we take the ratios between the posterior probabilities for different models.
Let $M_i$ and $M_j$ be the two models which we need to compare, then using Bayes theorem the ratio between their posterior probability $O_{ij}$ can be written as,
\begin{equation}
\label{ratio}
O_{ij}=\frac{p(M_i|D,I)}{p(M_j|D,I)}=\frac{p(M_i|I)p(D|M_i,I)}{p(M_j|I)p(D|M_j,I)}
\end{equation}  
Since $p(D|M_i,I)$ for the data $D$ is the likelihood for the model $M_i$, we re-notate it with $L(M_i)$, then the equation (\ref{ratio}) becomes,
\begin{equation}
O_{ij}=\frac{p(M_i|I)L(M_i)}{p(M_j|I)L(M_j)}.
\end{equation}
If the background information $I$ does not give any preference to a model over any other, then the prior probabilities becomes equal, so that,
\begin{equation}
\label{bayes}
O_{ij}=\frac{L(M_i)}{L(M_j)}\equiv B_{ij}
\end{equation}
where $B_{ij}$ is called the Bayes factor and is thus the ratio of the likelihood of the two models. This factor helps to compare two models with reference to their power in predicting the given data, hence it can be taken as a summary of the evidence provided by the data in favor of one model over the other \cite{kass}. If $1<B_{ij}<3$, then the model $M_i$ is not worth more than a bare mention. If $3<B_{ij}<20$, the strength of evidence of the model $M_i$ is positive. If $20<B_{ij}<150$, the evidence is strong and if $B_{ij}>150$, it is very strong \cite{Drell}. 

For a model having one or more free parameters such as $\alpha, \beta$,...etc, it's likelihood $L(M_i)$ can be evaluated as
\begin{equation}
L(M_i)=\int d\alpha \int d\beta....p(\alpha,\beta,...|M_i)L_i(\alpha,\beta,...),
\end{equation}
where $p(\alpha,\beta,...|M_i)$ is the prior probability for the set of parameter values $\alpha, \beta,...$  for the model $M_i$ to be true and $L_i(\alpha,\beta,...)$ is the likelihood for the combination of the parameters in the model and is usually taken as \cite{Drell},
\begin{equation}
L_i(\alpha,\beta,...)=exp[-\chi^2_i(\alpha,\beta,...)/2]
\end{equation}
where $\chi^2_i(\alpha,\beta,...)$ is the conventional $\chi^2$-function. If we assume that the model $M_i$ has two parameters $\alpha$ and $\beta$ having flat prior probabilities in some range, $[\alpha,\alpha+\Delta\alpha]$ and $[\beta,\beta+\Delta\beta]$, respectively, then the likelihood of the model is given as
\begin{equation}
\label{likelihood}
L(M_i)=\frac{1}{\Delta\alpha}\int_{\alpha}^{\alpha+\Delta\alpha}d\alpha L_i(\alpha),
\end{equation}
where 
\begin{equation}
L_i(\alpha)=\frac{1}{\Delta\beta}\int_{\beta}^{\beta+\Delta\beta}d\beta\, exp[-\chi^2(\alpha,\beta)/2]
\end{equation}
is called the marginal likelihood for the parameter $\alpha$. Similarly one can also find the marginal likelihood for the parameter $\beta$. This can be extended with the number of parameters. For instance, if we have three parameters, $\alpha, \beta, \gamma$, then the marginal likelihood of a parameter, say, $\alpha$, can be evaluated as,
\begin{equation}
\label{likelihood3}
L_i(\alpha)=\frac{1}{\Delta\gamma}\frac{1}{\Delta\beta }\int_{\gamma}^{\gamma+\Delta\gamma}d\gamma\int_{\beta}^{\beta+\Delta\beta} d\beta\, exp[-\chi^2(\alpha,\beta,\gamma)/2].
\end{equation} 

\section{Bulk viscous FLRW Universe }
\label{sec:flrw universe}
We can now proceed towards the Bayesian comparison of the different bulk viscous models of the universe.
First, we briefly discuss the bulk viscous model based on the reference \cite{Athira1}. The basic model we consider is a spatially flat universe, dominated with bulk viscous matter, described by the Friedmann equations,
\begin{equation}
\label{friedmann}\left(\frac{\dot{a}}{a}\right)^2=\frac{\rho}{3}
\end{equation}
\begin{equation}
\label{friedmanna}
2\frac{\ddot{a}}{a}+\left(\frac{\dot{a}}{a}\right)^{2}=-P^{*}
\end{equation}
where we have taken $8\pi G = 1$, $\rho$ is the density of the matter 
component of the universe, $a(t)$ is the scale factor  and an overdot represents the derivative with
respect to cosmic time $t$. We are neglecting the radiation components as it doesn't have any decisive role in the late evolution of the universe. The conservation equation of the cosmic component is then,
\begin{equation}
\label{conser} \dot{\rho}+3H\left(\rho+P^{*}\right)=0.
\end{equation}

According to Eckart theory \cite{Eckart1}, the effective pressure of the bulk
viscous fluid is given as
\begin{equation}
\label{p} P^{*}=P-3\zeta H
\end{equation}
where $P$ is the normal kinetic pressure and $\zeta$ is the
coefficient of bulk viscosity. The matter component in the late universe is non-relativistic hence it is usually taken as, $P=0.$ Then the contribution to
effective pressure is only due to the negative viscous pressure. The coefficient $\zeta$ is basically a transport coefficient, hence it would depend on the dynamics of the cosmic fluid. We consider the most general form for the bulk viscous coefficient $\zeta$, which is a linear combination of the three terms 
\cite{Athira1,Avelino,Athira2,ren1,Singh},
\begin{equation}
\label{zeta}
\zeta=\zeta_{0}+\zeta_{1}\frac{\dot{a}}{a}+\zeta_{2}\frac{\ddot{a}}{\dot{a}}
%=\zeta_{0}+\zeta_{1}H+\zeta_{2}(\frac{\dot{H}}{H}+H).
\end{equation} 
the first term $\zeta_{0}$, is a constant, the second is proportional to the velocity
and the third is proportional to the acceleration in the expansion of the universe.

Using the Friedmann equations and conservation equation, we can obtain the expression for the Hubble parameter $H$ for the general form of $\zeta$ as\cite{Athira1},
\begin{equation}
\label{hubbleina}
H(a)=H_{0}\left[a^{\frac{\tilde{\zeta}_{1}+\tilde{\zeta}_{2}-3}{2-\tilde{\zeta}_{2}}}\left(1+\frac{\tilde{\zeta}_{0}}{\tilde{\zeta}_{1}+\tilde{\zeta}_{2}-3}\right)-\frac{\tilde{\zeta}_{0}}{\tilde{\zeta}_{1}+\tilde{\zeta}_{2}-3}\right]
\end{equation}
where we define the dimensionless bulk viscous parameters
$\displaystyle 
 \tilde{\zeta}_{0}=\frac{3\zeta_{0}}{H_{0}},
\tilde{\zeta}_{1}=3\zeta_{1}$ and $ \tilde{\zeta}_{2}=3\zeta_{2}$.
The above equation can be integrated to obtain the scale factor as
\begin{equation}
\label{scalefactor}
a(t)=\left[\left(\frac{\tilde{\zeta}_{0}+\tilde{\zeta}_{12}-3}{\tilde{\zeta}_{0}}\right)+\left(\frac{3-\tilde{\zeta}_{12}}{\tilde{\zeta}_{0}}\right)
e^{\frac{\tilde{\zeta}_{0}}{2-\tilde{\zeta}_{2}}H_{0}(t-t_{0})}\right]^{\frac{2-\tilde{\zeta}_{2}}{3-\tilde{\zeta}_{12}}},
\end{equation} 
where $t_0$ is the present cosmic time and $\tilde{\zeta}_{12}=\tilde{\zeta}_{1}+\tilde{\zeta}_{2}$ .
For the best estimates of the parameters, it was shown in reference \cite{Athira1} that, in the limit 
$t-t_0 \to \infty$ the scale factor tends to $a(t)\to
e^{\frac{\tilde{\zeta}_{0}}{2-\tilde{\zeta}_{2}}H_{0}(t-t_{0})}$, corresponding to the de Sitter phase, while in the limit 
$t-t_0 \to 0$ the scale factor tends to $a(t)\to\left[1+\frac{3-\tilde{\zeta}_{12}}{2-\tilde{\zeta}_{2}}H_{0}(t-t_{0})\right]^{\frac{2-\tilde{\zeta}_{2}}{3-\tilde{\zeta}_{12}}}$, corresponding to an early decelerated expansion. Thus from the evolution of the scale factor, it is clear that the universe undergoes transition from the deceleration phase to accelerated phase and the transition redshift is found to be around 0.49. The model also predicts the present deceleration parameter around -0.68. However the age predicted was relatively low.

In reference \cite{Athira2}, we have done the phase space analysis of the model with bulk viscous matter as the dominating cosmic component. Three choices for the bulk viscous coefficient, (i)$\zeta=\zeta_{0}+\zeta_{1}\frac{\dot{a}}{a}+\zeta_{2}\frac{\ddot{a}}{\dot{a}}$, (ii)$\zeta=\zeta_{0}+\zeta_{1}\frac{\dot{a}}{a}$ and (iii)$\zeta=\zeta_{0}$ have been adopted. It was found that for all the three cases, it predicts a prior
unstable decelerated epoch, and a later stable
accelerating epoch, similar to the de Sitter phase. However, when the radiation component is also taken into account, the model support the conventional evolution of the universe, only for the case $\zeta=\zeta_{0}$. The other two cases doesn't predicts a
prior radiation dominated phase and conventional decelerated matter dominated phase of
the universe respectively.

\section{Bayesian Analysis of Bulk viscous models}
\label{sec:Analysis}
A fairly detailed description of the method of Bayesian analysis was given in section \ref{sec:bayes}. In this section we are going into the Bayesian analysis of the different bulk viscous models.
For this we consider the following cases separately,
\begin{enumerate}
\item 
$\zeta=\zeta_{0}+\zeta_{1}\frac{\dot{a}}{a}+\zeta_{2}\frac{\ddot{a}}{\dot{a}}$, \\ where viscosity is depending on both the velocity and acceleration of the expansion of
the universe.
\item
$\zeta=\zeta_{0}+\zeta_{1}\frac{\dot{a}}{a}$, \\ where viscosity is depending only on
velocity of the expansion of the universe apart from a constant additive part $\zeta_0$. 
This is equivalent to $\zeta=\zeta_{0}+\zeta_1\rho^s$, with $s=1/2$. 
\item  $\zeta=\zeta_{0}$, \\where viscosity is pure a constant
\item  $\zeta=\zeta_{1}\frac{\dot{a}}{a}$. \\ where viscosity only has the velocity dependent term and 
is equivalent to $\zeta=\zeta_1\rho^s$, with $s=1/2$. 
\item   $\zeta=\zeta_{0}+\zeta_{2}\frac{\ddot{a}}{\dot{a}}$ \\ where viscosity is depending on acceleration apart from an 
additive constant.
\end{enumerate}
The best estimated values of the parameters $(\tilde{\zeta}_0,\tilde{\zeta}_1,\tilde{\zeta}_2)$ corresponding to the cases 1, 2 and 3 are extracted in the reference \cite{Athira1,Athira2} and that of the  cases 4 and 5 are extracted using the same set of data and are given in the table \ref{tab:1}. The data used is the SCP
``Union"  Type Ia Supernova data ~\cite{kowalski1} composed of 307
data points from 13 independent data sets and the method used is $\chi^2$ minimization technique. 
\begin{table}[h]
\caption{Best estimates of the bulk viscous parameters, $H_{0}$ and also $\chi^{2}$ minimum value corresponding to the cases 4 and 5 of $\zeta$.
$\chi^{2}_{d.o.f}=\frac{\chi^{2}_{min}}{n-m}$, where $n=307$, the
number of data and $m$ is the number of parameters in the model. The
subscript d.o.f stands for degrees of freedom. For the best
estimation we have use SCP ``Union" 307 SNe Ia data sets. The values
of parameter corresponding to the first three cases are extracted in
~\cite{Athira1,Athira2}}
\label{tab:1}
\begin{tabular}{lll}
\hline\noalign{\smallskip}
Parameters & \multicolumn{2}{c}{Bulk viscous models}\\
{} & $\zeta=\zeta_{1}\frac{\dot{a}}{a}$ & $\zeta=\zeta_{0}+\zeta_{2}\frac{\ddot{a}}{\dot{a}}$\\
\noalign{\smallskip}\hline\noalign{\smallskip}
$\tilde{\zeta}_0$ & -- & 1.275\\
$\tilde{\zeta}_1$ & 1.683 & --\\
$\tilde{\zeta}_2$ & -- & 1.593 \\
$H_0$ & 69.21 & 70.50 \\
$\chi^2_{min}$ & 319.31 & 310.54 \\
$\chi^2_{d.o.f}$ & 1.04 & 1.02 \\
\noalign{\smallskip}\hline
\end{tabular} 
\end{table}

\begin{figure*}[h]
\centering
%\hspace{-1cm}%
\begin{minipage}[t]{.325\linewidth}
\includegraphics[width=1\textwidth]{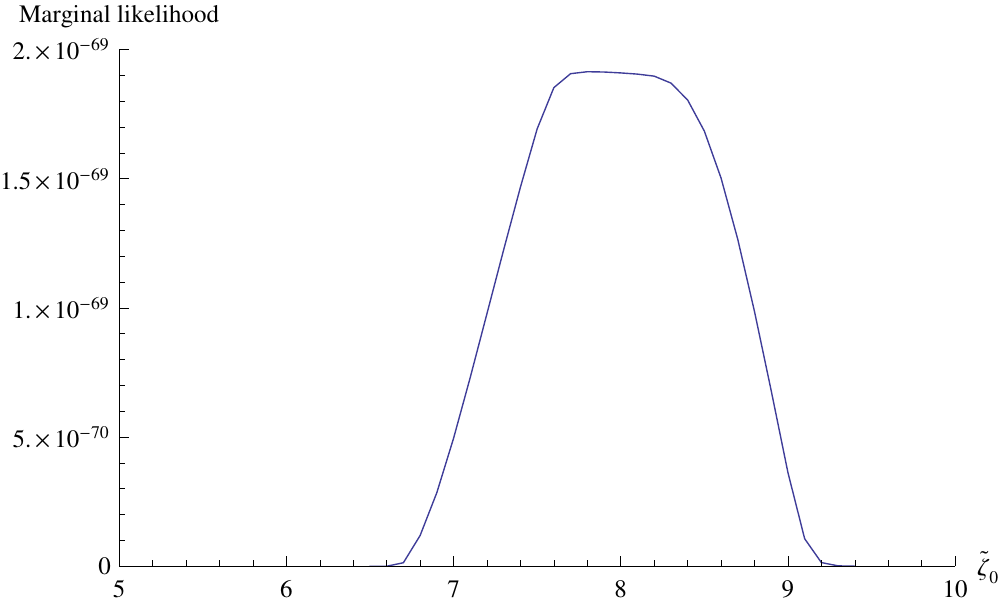}
\centering
%\subcaption{Marginal likelihood of $\tilde{\zeta}_0$ for case 1}
\end{minipage}%
\begin{minipage}[t]{.325\linewidth}
\centering
\includegraphics[width=1\textwidth]{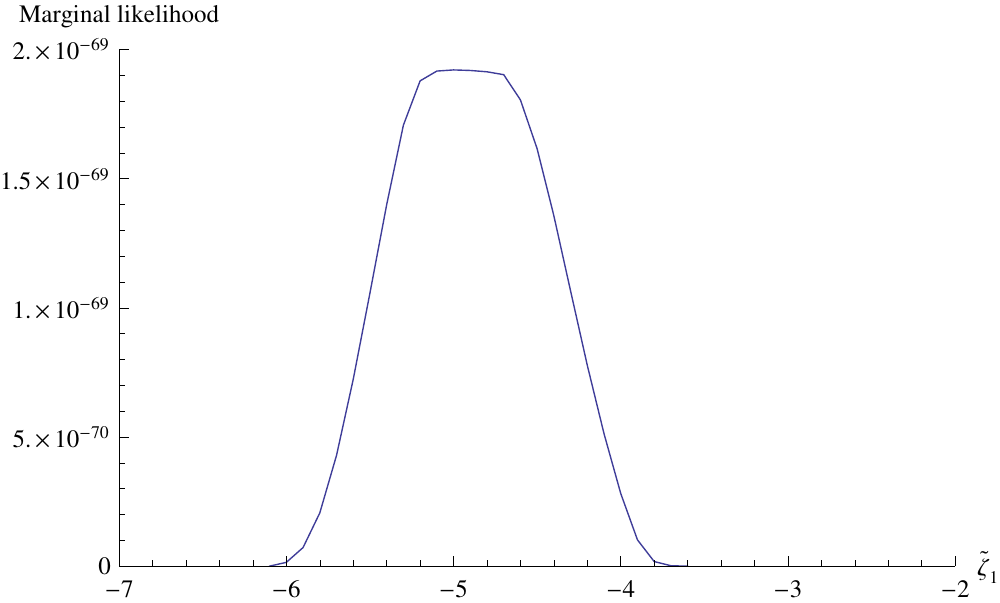}
%\subcaption{Marginal likelihood of $\tilde{\zeta}_1$ for case 1}
\end{minipage}
\begin{minipage}[t]{.325\linewidth}
\centering
\includegraphics[width=1\textwidth]{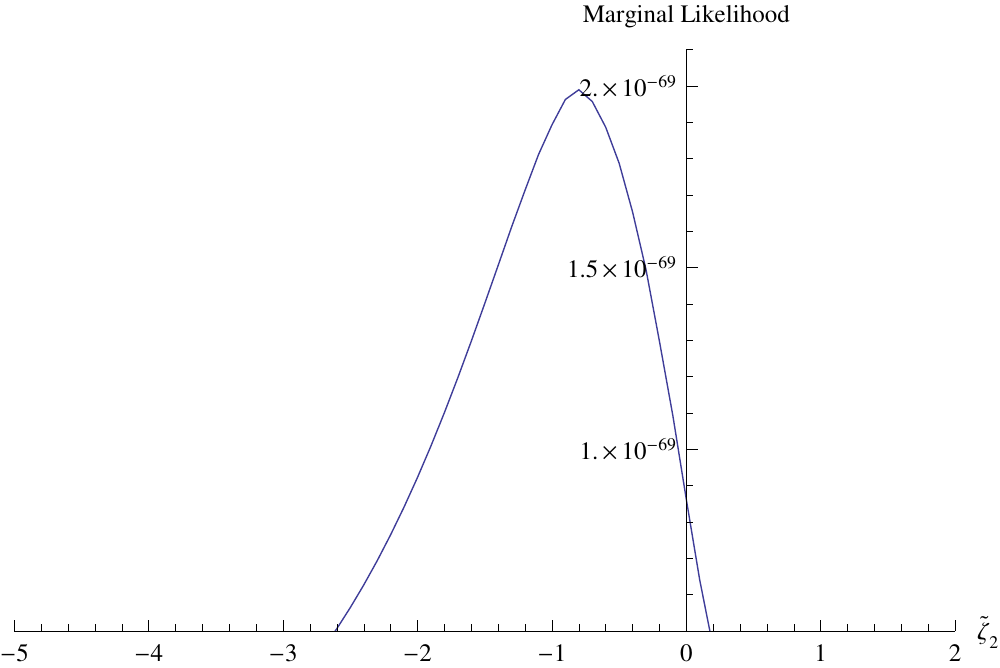}
%\subcaption{Marginal likelihood of $\tilde{\zeta}_2$ for case 1}
\end{minipage}
\caption{Marginal Likelihood of the parameters $\tilde{\zeta}_0$, $\tilde{\zeta}_1$ and $\tilde{\zeta}_2$ corresponding to the case 1, when $\zeta=\zeta_{0}+\zeta_{1}\frac{\dot{a}}{a}+\zeta_{2}\frac{\ddot{a}}{\dot{a}}$.\label{fig:case 1}}
\end{figure*}

\begin{figure*}[h]
\centering
%\hspace{-1cm}%
\begin{minipage}[t]{.5\linewidth}
\includegraphics[width=.65\textwidth]{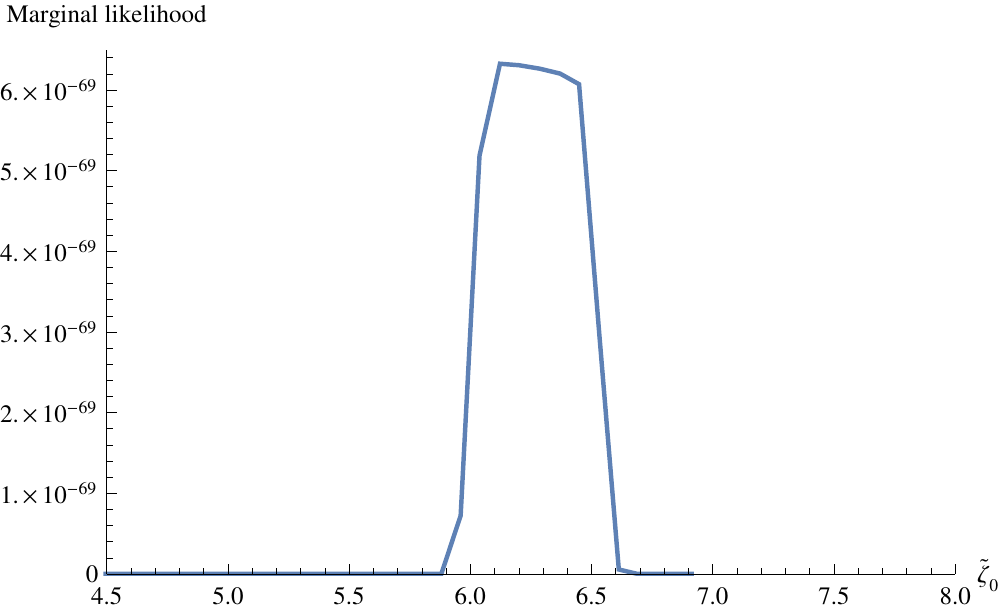}
\centering
%\subcaption{Marginal likelihood of $\tilde{\zeta}_0$ for case 1}
\end{minipage}%
\begin{minipage}[t]{.5\linewidth}
\centering
\includegraphics[width=.65\textwidth]{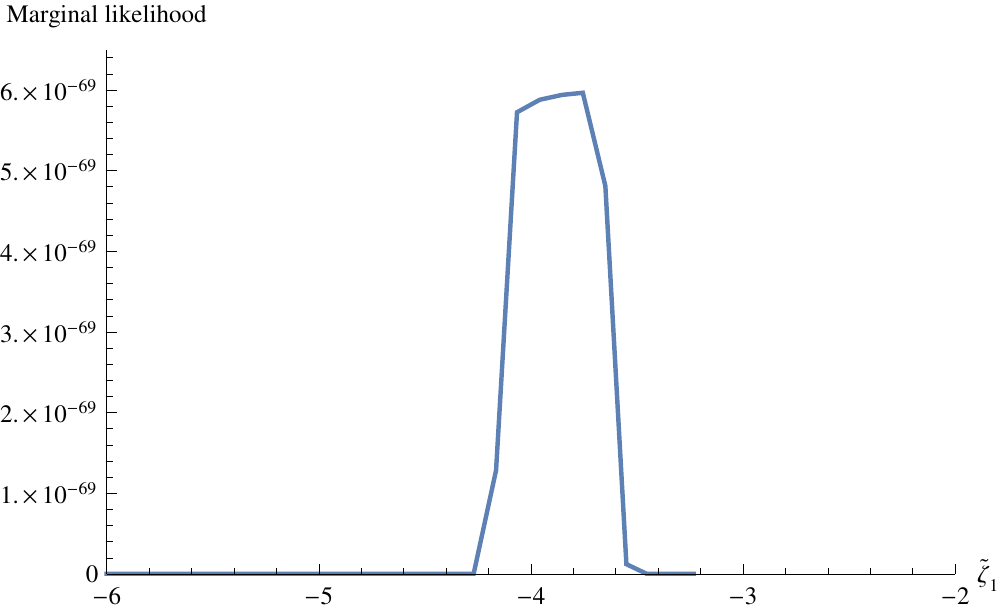}
%\subcaption{Marginal likelihood of $\tilde{\zeta}_1$ for case 1}
\end{minipage}
\caption{Marginal Likelihood of the parameters $\tilde{\zeta}_0$ and $\tilde{\zeta}_1$ corresponding to the case 2, when $\zeta=\zeta_{0}+\zeta_{1}\frac{\dot{a}}{a}$.\label{fig:case 2}}
\end{figure*}

\begin{figure}[h]
\centering
%\hspace{-1cm}%
\includegraphics[width=.4\textwidth]{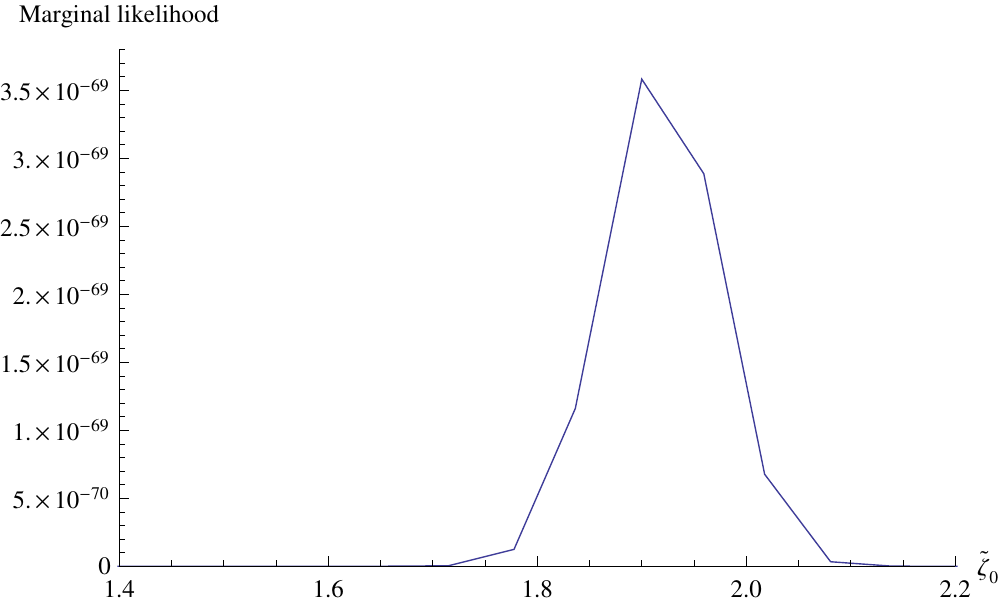}
\centering
%\subcaption{Marginal likelihood of $\tilde{\zeta}_0$ for case 1}
\caption{Marginal Likelihood of the parameter $\tilde{\zeta}_0$ corresponding to the case 3, when $\zeta=\zeta_{0}$, a constant.\label{fig:case 3}}
\end{figure}

\begin{figure}[h]
\centering
%\hspace{-1cm}%
\includegraphics[width=.4\textwidth]{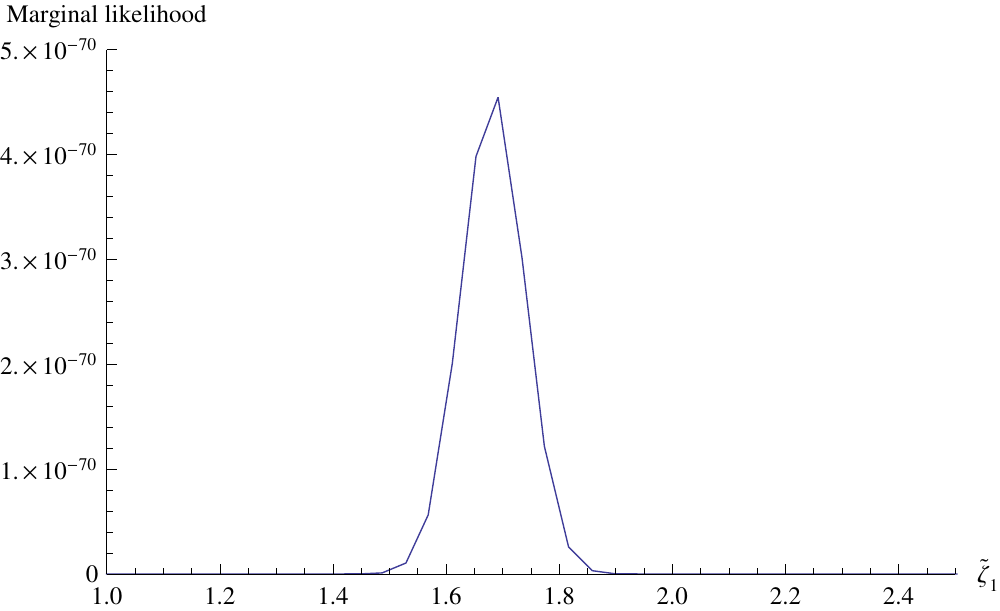}
\centering
%\subcaption{Marginal likelihood of $\tilde{\zeta}_0$ for case 1}
\caption{Marginal Likelihood of the parameter $\tilde{\zeta}_1$ corresponding to the case 4, when $\zeta=\zeta_{1}\frac{\dot{a}}{a}$, a constant.\label{fig:case 4}}
\end{figure}

\begin{figure*}[h]
\centering
%\hspace{-1cm}%
\begin{minipage}[t]{.5\linewidth}
\includegraphics[width=.65\textwidth]{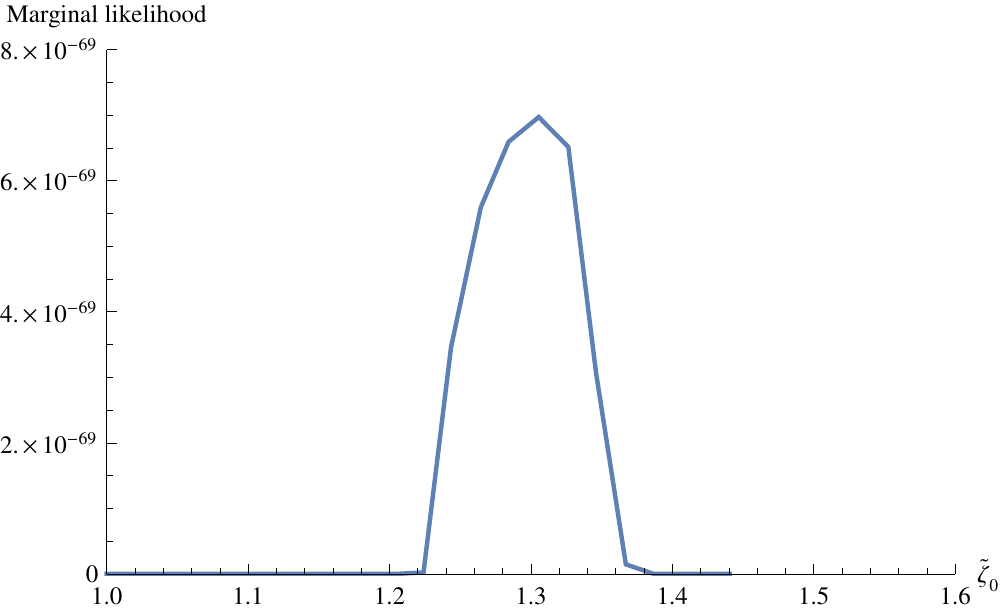}
\centering
%\subcaption{Marginal likelihood of $\tilde{\zeta}_0$ for case 1}
\end{minipage}%
\begin{minipage}[t]{.5\linewidth}
\centering
\includegraphics[width=.65\textwidth]{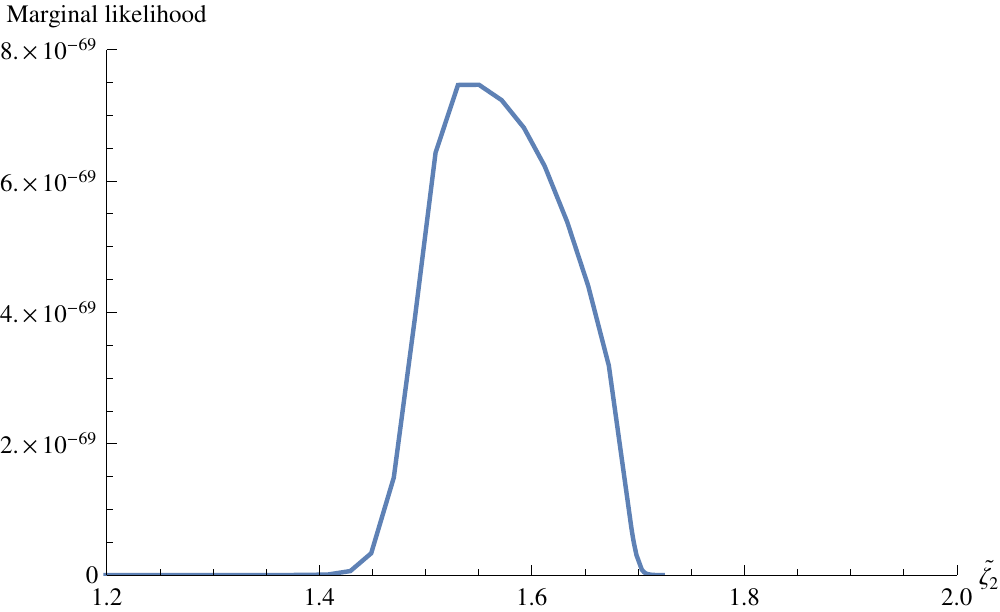}
%\subcaption{Marginal likelihood of $\tilde{\zeta}_1$ for case 1}
\end{minipage}
\caption{Marginal Likelihood of the parameters $\tilde{\zeta}_0$ and $\tilde{\zeta}_2$ corresponding to the case 5, when $\zeta=\zeta_{0}+\zeta_{2}\frac{\ddot{a}}{\dot{a}}$.\label{fig:case 5}}
\end{figure*}

The $\chi^{2}$ function is constructed as,
\begin{equation}
\label{chisquare}
\chi^{2}\equiv
\sum^{n}_{k=1}\frac{\left[\mu_{t}-\mu_{k}\right]^{2}}{\sigma_{k}^{2}},
\end{equation}
where $\mu_{k}$ is the observational distance modulus for the k-th
Supernova (obtained from the data) with red shift $z_k$, $\sigma_{k}^{2}$ is the variance of the measurement,
$n$ is the total number of data and $\mu_{t}$ is the theoretical distance modulus for the
k-th Supernova with the same redshift $z_{k}$, which is given as
\begin{equation}
\mu_{t}=m-M=5\log_{10}[\frac{d_{L}}{Mpc}]+25
\end{equation}
where, $m$ and $M$ are the apparent and absolute magnitudes of the
SNe respectively. $d_{L}$ is the luminosity distance and is defined as
\begin{equation}
d_{L}=c(1+z)\int_{0}^{z}\frac{dz'}{H},
\end{equation}
where $c$ is the speed of light. After obtaining the $\chi^2,$ we evaluate the marginal likelihood, using equation (\ref{likelihood3}), and likelihood, using equation (\ref{likelihood}), for all the five cases of the model. We kept $\Lambda$CDM model as the reference model in order to compare the bulk viscous models and calculate the Bayes factor using equation (\ref{bayes}). The marginal likelihood of the parameters $\tilde{\zeta}$ corresponding to the five cases of bulk viscous models are shown in figures \ref{fig:case 1}, \ref{fig:case 2}, \ref{fig:case 3}, \ref{fig:case 4} and \ref{fig:case 5}, respectively.


We obtained the Bayes factor for three different priors for the parameters $\tilde{\zeta}_0, \, \tilde{\zeta}_1, \, \tilde{\zeta}_2$. Since there is no prior information regarding any of the viscous parameters, we assume flat prior for all the three parameters.. Each prior corresponds to a range of values of  $\tilde\zeta$'s for a fixed likelihood. For choosing the prior for $\tilde{\zeta}$'s we have plotted $exp(-\chi^2 / 2)$ as a function of a given $\tilde{\zeta}$ i.e., $\tilde{\zeta}_{0},\tilde{\zeta}_1$, or $\tilde{\zeta}_2$, by keeping the other two as constants (equal to that corresponding to the minimum of $\chi^2$).  For instance prior I, corresponds to the range of $\tilde\zeta$'s for likelihood of about $1*10^{-70},$ prior II corresponds to the range of $\tilde\zeta$'s for likelihood of about $1*10^{-80}$ and prior III corresponds to the range of $\tilde\zeta$'s for likelihood of about $1*10^{-90}$. The results are tabulated in the Table \ref{tab:3}. It can be seen from the table that for the first two cases the Bayes factor depends on the prior, while for the remaining three case such strong dependence on the priors are not evident. For case 2, i.e., when $\zeta=\zeta_0+\zeta_{1}\frac{\dot{a}}{a}$, there is an increase in Bayes factor with prior and it exceeds 3, giving more evidence for its increasing strength. The important thing to be noted here is about the value of the Bayes factor for the respective model. For the cases $\zeta=\zeta_{0}$ and $\zeta=\zeta_{1}\frac{\dot{a}}{a},$ the factor is much less than one. While for other cases the values are above one. As per the standard classification, it can be mentioned that, the models for which the Bayes factor is in between 1 and 3, can have only a very feeble advantage over the standard $\Lambda$CDM model, however is not worth more than a bare mention. 
\begin{table}[h]
\caption{Bayes factors with respect to $\Lambda$CDM model corresponding to three different priors}
\label{tab:3}
\begin{tabular}{lllll}
\hline\noalign{\smallskip}
Sl. No. & Bulk viscous models &  \multicolumn{3}{c}{Bayes factor $B_{i\Lambda}=\frac{L(M_i)}{L(M_\Lambda)}$} \\
{} & $M_i$ & Prior I & Prior II & Prior III\\
\noalign{\smallskip}\hline\noalign{\smallskip}
1 & $\zeta_{0}+\zeta_{1}\frac{\dot{a}}{a}+\zeta_{2}\frac{\ddot{a}}{\dot{a}}$ & 0.743 & 1.372 & 1.043 \\
2 & $\zeta_{0}+\zeta_{1}\frac{\dot{a}}{a}$ & 1.86 & 2.63 & 3.63 \\
3 & $\zeta_{0}$ & 0.27 & 0.32 & 0.42 \\
4 & $\zeta_{1}\frac{\dot{a}}{a}$ & 0.05 & 0.042 & 0.052 \\
5 & $\zeta_{0}+\zeta_{2}\frac{\ddot{a}}{\dot{a}}$ & 1.65 & 1.91 & 1.77 \\
\noalign{\smallskip}\hline
\end{tabular}
\end{table}

The SNe data that we have used contains the magnitude of supernovae in the red-shift range $0.015<z<1.55$.
The data predicts a transition from an early decelerated epoch to the late acceleration at a redshift of around $z\sim0.5$ \cite{Athira1}. Among the full data set, the low redshift data within the range $0.015<z<0.5$ is often used for deducing the current value of the Hubble parameter. 
The high redshift data, which were obtained with small interference with the background and with high accuracy, are considered to be the best part of the data. For a critical analysis we have repeated our computation using the high redshift part of the data, corresponding to the red shift range, $0.5<z<1.55$. For this range, we have extracted the values of the parameters, $\tilde{\zeta}$'s corresponding to the above 5 cases using the $\chi^2$ minimization technique. The results are tabulated in table \ref{tab:2}.
\begin{table}[h]
\caption{Best estimates of the Bulk viscous parameters, $H_{0}$ and also $\chi^{2}$ minimum value corresponding to the different cases of $\zeta$ for high redshift.
$\chi^{2}_{d.o.f}=\frac{\chi^{2}_{min}}{n-m}$, where $n=150$, the
number of data and $m$ is the number of parameters in the model. The
subscript d.o.f stands for degrees of freedom.}
\label{tab:2}
\begin{tabular}{lllllll}
\hline\noalign{\smallskip}
Viscous models & $\tilde{\zeta}_0$ & $\tilde{\zeta}_1$ & $\tilde{\zeta}_2$ & $H_0$ & $\chi^2_{min}$ & $\chi^2_{d.o.f}$  \\
\noalign{\smallskip}\hline\noalign{\smallskip}
$\zeta_{0}+\zeta_{1}\frac{\dot{a}}{a}+\zeta_{2}\frac{\ddot{a}}{\dot{a}}$ & 7.36 & -4.73 & -1.25 & 67.41 & 166.88 & 1.135 \\
$\zeta_{0}+\zeta_{1}\frac{\dot{a}}{a}$ & 4.53 & -2.53 & -- & 67.41 & 166.88 & 1.13 \\ 
$\zeta_{0}$ & 1.17 & -- & -- & 63.06 & 167.01 & 1.12\\
$\zeta_{1}\frac{\dot{a}}{a}$ & -- & 0.787 & -- & 61.43 & 167.09 & 1.12  \\
$\zeta_{0}+\zeta_{2}\frac{\ddot{a}}{\dot{a}}$ & 1.28 & -- & 1.43 & 67.41 & 166.88 & 1.13 \\
\noalign{\smallskip}\hline
\end{tabular} 
\end{table}

The Bayes factor of the bulk viscous models corresponding to the five cases for high redshift data are tabulated in the Table \ref{tab:4}. 
\begin{table}[h]
\caption{Bayes factors with respect to $\Lambda$CDM model corresponding to two different priors for high redshift.}
\label{tab:4}
\begin{tabular}{llll}
\hline\noalign{\smallskip}
Sl. No. & Bulk viscous models &  \multicolumn{2}{c}{Bayes factor $B_{i\Lambda}=\frac{L(M_i)}{L(M_\Lambda)}$} \\
{} & $M_i$ & Prior I & Prior II \\
\noalign{\smallskip}\hline\noalign{\smallskip}
1 & $\zeta_{0}+\zeta_{1}\frac{\dot{a}}{a}+\zeta_{2}\frac{\ddot{a}}{\dot{a}}$ & 0.79 & 0.06 \\
2 & $\zeta_{0}+\zeta_{1}\frac{\dot{a}}{a}$ & 1.044 & 1.4 \\
3 & $\zeta_{0}$ & 1.17 & 1.48 \\
4 & $\zeta_{1}\frac{\dot{a}}{a}$ & 1.13 & 1.42 \\
5 & $\zeta_{0}+\zeta_{2}\frac{\ddot{a}}{\dot{a}}$ & 0.733 & 0.2685\\
\noalign{\smallskip}\hline
\end{tabular}
\end{table}
Here prior I corresponds to the range of $\tilde\zeta$'s for likelihood of about $1*10^{-38}$ and prior II corresponds to the range of $\tilde\zeta$'s 
for likelihood of about $1*10^{-45}$. The models corresponding to the cases 1 and 5, where the viscous coefficient depends on the acceleration of the 
expansion, have Bayes factor less than one for both the priors. As a result these two cases are not worth of any mention against the standard model.  
This indicates that the dependence of viscosity on the acceleration is not so sensitive. Riess et. al. have found that the magnitude of acceleration is
small, since the distance of the high redshift supernovae were on average only $10\%-15\%$ farther than expected in a universe with mass density parameter
$\Omega_m\sim 0.3.$ Such a small acceleration would not have any observable effect on the transport coefficients like that of viscosity. For cases 2, 3 and 
4, the Bayes factors are larger than one and it increases slightly with prior. Among these, the third case is of constant viscosity, while for second and 
fourth cases, the viscous coefficient depends on the velocity of expansion of the universe. As seen from the Table \ref{tab:4}, the Bayes factors for the 
cases 2, 3 and 4 are all in the range $1<B_{ij}<3$ and it seems quite difficult to discriminate between them. All these three cases are thus qualified to 
have bare mention against the the $\Lambda$CDM model.

\section{Conclusion}
\label{Sec:Conclusion}
Bulk viscous models of the universe are important, especially regarding the unification of dark 
matter and dark energy. In some earlier works we have shown that this class of models reasonably explained the background evolution of the universe. It was also shown from the dynamical behavior that the model   
describes all the conventional phases of the universe and asymptotically tend towards a stable de Sitter epoch, provided the bulk viscosity of the cosmic fluid is a constant. In the present work we have contrasted the bulk viscous model of the universe with the standard $\Lambda$CDM model using the method of Bayesian analysis. We have first extracted the viscous parameters corresponding to the following five cases of bulk viscous model, (1)
$\zeta=\zeta_{0}+\zeta_{1}\frac{\dot{a}}{a}+\zeta_{2}\frac{\ddot{a}}{\dot{a}}$, (2)
$\zeta=\zeta_{0}+\zeta_{1}\frac{\dot{a}}{a}$, (3) $\zeta=\zeta_{0}$, (4) $\zeta=\zeta_{1}\frac{\dot{a}}{a}$, (5) $\zeta=\zeta_{0}+\zeta_{2}\frac{\ddot{a}}{\dot{a}}$, using the ``Union" data of Supernovae type Ia. We have obtained the Bayes factor for all the five cases, see table \ref{tab:3}. For the full supernovae data set, the results indicate that the model corresponding to case 2, i.e., $\zeta=\zeta_0+\zeta_1 \frac{\dot a}{a}$ have a Bayes factor just above 3, and thus have slight advantage over the $\Lambda$CDM model compared with other cases. For the model corresponding to cases 1 and 5 , the Bayes factor is just above one and can just have a bare mention, in contrast to the standard model. All other cases, especially the case 3, with constant viscosity, seems to fail in standing against the standard model.  

For more reliable result, we restrict to supernovae data with relatively high redshifts, $z>0.5$,  
which were obtained with less background interference and hence are more reliable in making predictions regarding the evolution near the transition. The results consequent to this have a marked deviation from the previous one, such that the Bayes factor for the constant bulk viscosity $\zeta=\zeta_0$  (case 3) and models corresponding to cases 2 and 4, where the viscous coefficient depends on the velocity of expansion, are having a slight advantage over other cases when compared with the standard $\Lambda$CDM model. Since  Bayes factors of the cases 2, 3 and 4, are all in the range $1<B_{ij}<3,$ it is difficult to discriminate    
among themselves. However it was shown in reference \cite{Athira2} that only the case 3 will have asymptotically stable end de Sitter phase. Taking account of this, it can be concluded that, among the cases 2, 3 and 4, which are having almost same Bayes factor, the case 3, corresponding to the one with stable end de Sitter phase, can be preferred over the other cases.

\end{document}